%% file: main_RB.tex
\documentclass[letterpaper, 10 pt, conference]{ieeeconf}  % Comment this line out if you need a4paper

\IEEEoverridecommandlockouts                              % This command is only needed if 
\usepackage{graphics} % for pdf, bitmapped graphics files
\usepackage{epsfig} % for postscript graphics files
\usepackage{times} % assumes new font selection scheme installed
\usepackage{amsmath} % assumes amsmath package installed
\usepackage{amssymb}  % assumes amsmath package installed
\usepackage{color}
\usepackage{cite}
\usepackage{algorithm}
\usepackage{algorithmicx}
\usepackage{algpseudocode}
\usepackage{booktabs}
\usepackage{url}

\newtheorem{remark}{Remark}

\title{\LARGE \bf
In-Context Learning for Zero-Shot Speed Estimation of BLDC motors
}

\author{Alessandro Colombo$^{1}$, Riccardo Busetto$^{2}$, Valentina Breschi$^{3}$,\\Marco Forgione$^{2}$, Dario Piga$^{2}$, and Simone Formentin$^{1}$% <-this % stops a space
\thanks{$^{1}$Dip. di Elettronica, Informazione e Bioingegneria, Politecnico di Milano, Milano, IT
        {\tt\small {alessandro4.colombo, simone.formentin}@polimi.it}}%
\thanks{$^{2}$IDSIA Dalle Molle Institute for Artificial
Intelligence, SUPSI, Lugano-Viganello, CH 
        {\tt\small {name}.{surname}@supsi.ch}}%
\thanks{$^{3}$Dept. of Electrical Engineering, Eindhoven University of Technology, Eindhoven, NL
        {\tt\small v.breschi@tue.nl}}%
}

\begin{document}

\maketitle
\thispagestyle{empty}
\pagestyle{empty}

%%%%%%%%%%%%%%%%%%%%%%%%%%%%%%%%%%%%%%%%%%%%%%%%%%%%%%%%%%%%%%%%%%%%%%%%%%%%%%%%
\begin{abstract}
Accurate speed estimation in sensorless brushless DC motors is essential for high-performance control and monitoring, yet conventional model-based approaches struggle with system nonlinearities and parameter uncertainties. 
In this work, we propose an in-context learning framework leveraging transformer-based models to perform zero-shot speed estimation using only electrical measurements. 
By training the filter offline on simulated motor trajectories, we enable real-time inference on unseen real motors without retraining, eliminating the need for explicit system identification while retaining adaptability to varying operating conditions.
Experimental results demonstrate that our method outperforms traditional Kalman filter-based estimators, especially in low-speed regimes that are crucial during motor startup.

% \emph{Keywords:} State estimation, in-context learning, electric drives.
\end{abstract}

%%%%%%%%%%%%%%%%%%%%%%%%%%%%%%%%%%%%%%%%%%%%%%%%%%%%%%%%%%%%%%%%%%%%%%%%%%%%%%%%
%%%%%%%%%%%%%%%%%%%%%%%%%%%%%%%%%%%%%%%%%%%%%%%%%%%%%%%%%%%%%%%%%%%%%%%%%%%%%%%%
\input{sections/introduction_RB}

\input{sections/problem_statement}

\input{sections/methodology}

\input{sections/experimental_results_AC}

\input{sections/conclusions}

\bibliographystyle{plain}
\bibliography{references}

\addtolength{\textheight}{-12cm}   % This command serves to balance the column lengths

%%%%%%%%%%%%%%%%%%%%%%%%%%%%%%%%%%%%%%%%%%%%%%%%%%%%%%%%%%%%%%%%%%%%%%%%%%%%%%%%

%%%%%%%%%%%%%%%%%%%%%%%%%%%%%%%%%%%%%%%%%%%%%%%%%%%%%%%%%%%%%%%%%%%%%%%%%%%%%%%%

%%%%%%%%%%%%%%%%%%%%%%%%%%%%%%%%%%%%%%%%%%%%%%%%%%%%%%%%%%%%%%%%%%%%%%%%%%%%%%%%

%%%%%%%%%%%%%%%%%%%%%%%%%%%%%%%%%%%%%%%%%%%%%%%%%%%%%%%%%%%%%%%%%%%%%%%%%%%%%%%%

\end{document}

%% file: sections/introduction_RB.tex
\section{introduction}\label{sec:introduction}
Enabled by advanced control strategies such as field-oriented control (FOC), nowadays \textit{Brushless DC} (BLDC) motors guarantee high dynamic performance, efficiency, and minimal maintenance requirements \cite{mohanraj2022review}. For these reasons, Brushless DC motors are widely used across various applications, including industrial automation, robotics, and household appliances. However, in many scenarios, cost constraints require sensorless control, eliminating direct rotor position and speed sensors \cite{akrami2024sensorless}. In such cases, estimation algorithms become essential to ensure precise motor control.
This is true in particular for speed estimation, as accurate speed estimates %estimation in sensorless BLDC motors is
are crucial to achieving high-performance operation and reliable control in sensorless BLDC. Given the importance of speed estimation, several estimation techniques thus exist to tackle this problem. Among the most common ones, we have back-EMF-based methods \cite{lai2008novel, attar2021control} and observer-based approaches \cite{chen2018position,lavanya2009sensorless}. The first set of approaches leverages the voltage induced by rotor motion to estimate speed but leads to poor estimates at low speeds due to low SNR. The latter methods, such as Kalman filters \cite{lavanya2009sensorless} and sliding-mode observers \cite{qiao2012new}, estimate motor states but are often sensitive to model uncertainties and parameter variations.

More recently, data-driven approaches based on machine learning and neural networks have emerged as promising alternatives, demonstrating improved robustness against model uncertainties \cite{del2014development}. These methods learn motor dynamics directly from data, enabling adaptive and generalizable estimators. Despite their advantages, data-driven methods require substantial training data from each specific system, making their deployment impractical in many real-world applications. From this key practical issue stems the need for tools that allow the transfer of knowledge from data coming from similar, yet not exactly equal, domains, e.g., data generated by a system and its digital twins (which represent the system of interest but cannot fully capture real-world complexities). 

Recent breakthroughs in deep learning and transformer-based architectures \cite{vaswani2017attention, radford2019language} have demonstrated remarkable capabilities in coping with these domain shifts in %as the domain shifts that occur among real systems and digital twins, which cannot fully capture real-world complexities. In this direction, the recent breakthroughs in deep learning and transformer-based architectures \cite{vaswani2017attention, radford2019language} have demonstrated remarkable capabilities in 
sequence modeling, including system identification \cite{du2023can, forgione2023system}. In particular, these works show that a transformer trained on multiple datasets representing different system instances can generalize to unseen systems without explicit re-training. This concept is extended to state estimation in \cite{goel2024can}, where it is demonstrated that a transformer can effectively approximate Kalman Filter formulations. Inspired by these developments, \cite{bus24} propose a novel \textit{in-context learning} (ICL) framework that leverages transformer-based architectures for zero-shot state estimation of hidden states in dynamical systems.  The core idea of the approach is to leverage the transformer’s ability to learn and generalize from sequential data, implicitly capturing the dynamics from a history of input-output measurements. Unlike conventional methods that require extensive system identification and parameter tuning (see, e.g., \cite{khan2025optimized}), this approach thus relies on a contextual filter trained offline on simulated input-output trajectories. Then, a%
t deployment, the trained filter automatically adapts to a new instance by using the input-output sequence as context, eliminating the need for fine-tuning. This allows for real-time inference on previously unseen systems without retraining, significantly improving adaptability to varying operating conditions and, hence, reducing deployment time.

The key contribution of this work is to validate this methodology in a real experimental case study, namely sensorless speed estimation in BLDC motors. This problem is an ideal testbed for assessing the practical feasibility of the approach, as several (practical) challenges must be addressed to transition from the methodology proposed in \cite{bus24} and its deployment in the considered real-world application.
For example, the in-context learning framework %requires 
relies on data gathered over multiple experiments. This requirement is clearly impractical when one wants to obtain these data by performing real-world experiments. We hence propose to use an out-of-the-box BLDC simulator to collect the training data, thus training the filter with a \textit{sim-to-real logic}. This allows us to show how to deal with distribution shifts between simulated and real-world trajectories, which was not explicitly considered in \cite{bus24}.
A second challenge is related to computational constraints that have to be faced in real-world applications. Indeed, the model must operate within the system’s sampling time while preserving real-time feasibility. This introduces additional non-idealities that have to be accounted for, such as aliasing at higher speeds, which can degrade estimator accuracy. In this work, we show how to cope with this issue by leveraging the (mechanical) characteristics of the motor in the training of the transformer.
To evaluate the proposed approach, the transformer-based contextual filter is tested on real, unseen BLDC motor instances. Our results show the method successfully generalizes across different motors that without requiring any parameter updates. Moreover, when compared against an \textit{Extended Kalman Filter} (EKF) baseline, the transformer-based estimator demonstrates superior performance, particularly in low-speed regimes, where conventional methods struggle due to weak back-EMF signals. This validates the findings of previous numerical studies and confirms the real-world applicability of in-context learning for motor state estimation.

The paper is organized as follows. In Section~\ref{sec:problem_statement}, we introduce the settings and the problem formulation. Details regarding the methodology are provided in Section~\ref{sec:methodology}, including further details on the model architecture and how to deal with aliasing. In Section~\ref{sec:experimental_results}, our experimental setup is presented along with the simulator employed for generating the training data. A brief description of the implemented EKF benchmark is then given before the analysis of the results. Section~\ref{sec:conclusions} ends the paper, with some concluding remarks and directions for future works.

%% file: sections/problem_statement.tex
\section{Problem statement}\label{sec:problem_statement}
Consider a function describing the class of \textit{brushless DC} (BLDC) motors $\mathcal{S}(\theta)$, where $\theta \in \mathbb{R}^{n_\theta}$ are parameters describing the nonlinear state dynamics of a motor instance.
Each motor $S^{(n)}=\mathcal{S}(\theta^{(n)})$ belonging to $\mathcal{S}$ is characterized by a discrete dynamics
% \begin{subequations}\label{eq:dynamics}
% \begin{alignat}{3}
%     x_k &= f_\theta(x_k,u_k) + w_{w,k}\\
%     y_k &= g(x_k,u_k) + w_{v,k},
% \end{alignat}
% \end{subequations}
% 
\begin{subequations}\label{eq:dynamics}
\begin{alignat}{3}
    x_{k+1} &= f_\theta(x_k,u_k) + \eta_k\\
    y_k &= g(x_k,u_k) + \xi_k,
\end{alignat}
\end{subequations}
where the state, input and output vectors are, respectively 
\begin{subequations}\label{eq:vectors}
\begin{align}
    \label{eq:state_vector}
    x_k &= \begin{bmatrix}i_{\alpha,k}, i_{\beta,k}, \omega_k, \theta_{e,k},
    \end{bmatrix}^{\top}\\  
    \label{eq:input_vector}
    u_k &= \begin{bmatrix}v_{\alpha,k},  v_{\beta,k}\end{bmatrix}^{\top},\\
    y_k &= \begin{bmatrix}i_{\alpha,k}, i_{\beta,k}\end{bmatrix}^{\top}.
\end{align}
\end{subequations}
Above, $v_\alpha$, $v_\beta$ [V] and $i_\alpha$, $i_\beta$ [A] are the voltages and currents projected on the $\alpha-\beta$ plane through the Clarke Transform, $\theta_{e,k}$ [rad] is the electrical angle of the rotor and $\omega_k$ [rad/s] the mechanical speed. $\eta_k$, $\xi_k$ are process and output noises affecting the system, assumed white with zero-mean.
Notice that the output dynamics is linear and describes a sensorless case where $\theta_{e,k}$ and $\omega_k$ are non-measured states. Finally the subscript $\theta$ remarks the dependency of the motor on the parameters that define the particular realization from the class $\mathcal{S}$.

The goal is to design a filter to estimate the motor speed $\omega_k$ of any system $S^{(n)}$ belonging to the class $\mathcal{S}$, from measurements available up to time instant $k$. 

To this end, we assume to have access to a meta-dataset 
\begin{equation}
    \label{eq:meta-set}
    \mathcal{D} = \{  \{ v^{(n)}_{\alpha,k}, v^{(n)}_{\beta,k}, i^{(n)}_{\alpha,k}, i^{(n)}_{\beta,k}, \omega^{(n)}_k\}^{T^{(n)}}_{k=1} \}_{n=1}^{b}
\end{equation}
containing trajectories of duration $T^{(n)}$ from $n=1,\ldots,b$ motor instances whose dynamics corresponds to values of $\theta$ sampled from the user-specified distribution $p(\theta)$, e.g., a box containing $\theta$. Similarly, inputs $u^{(n)}_k$ also belong to a prior distribution $p(u_k)$, that must be coherent with the one that describes the control variables in the application of interest for the end-user.
In practice, the trajectories in \eqref{eq:meta-set} are available performing an experiment of duration $T^{(n)}$ either with a \textit{black-box} simulator, or on a real motor of \textit{unknown} dynamics.

More specifically, we use a window of finite length $H \leq \min_i T^{(i)}$ of past and present inputs and measurements (information vector) from \eqref{eq:meta-set}
\begin{equation}\label{eq:information_vector}
    I^{(n)}_{k} = \{ v^{(n)}_{\alpha,\kappa}, v^{(n)}_{\beta,\kappa}, i^{(n)}_{\alpha,\kappa}, i^{(n)}_{\beta,\kappa} \}^{k}_{\kappa=k-H+1},
\end{equation}
as an input to a filter $\mathcal{F}_\phi$ such that
\begin{equation}
    \label{eq:contextual_filter}
    \hat{\omega}_{\phi,k}^{(n)} = \mathcal{F}_\phi(I^{(n)}_{k}),
\end{equation}
with $\hat{\omega}_{\phi,k}^{(n)}$ the reconstructed speed at time instant $k$ for any BLDC instance $i$ within the class $\mathcal{S}$, and $\phi$ design parameters of the filter to optimize such that the estimation cost
\begin{equation}\label{eq:cost}
    J(\phi) =\frac{1}{b\cdot H}\sum_{i=1}^{b}{\sum_{k=1}^{H}}{\| {\omega}_k^{(n)} -  \hat{\omega}_{\phi,k}^{(n)} \|}
\end{equation}
is minimized.

%% file: sections/methodology.tex
\section{Methodology}\label{sec:methodology}
% The filter in \eqref{eq:contextual_filter} estimates the motor speed at instant $k$ for any BLDC motor belonging to the class. Specifically, the filter generalizes to unseen motor instances without the need of parameter updates by leveraging \textit{in-context learning} (ICL). In this paradigm, underlying patterns --such as the underlying dynamics of the motor- are inferred from a provided context, i.e., a sequence of input-output measurements.

% In our application, the \textit{context} coincides with the sequence of input-ouput data $I_k^{(n)}$. Indeed, for different realizations of $\theta$, the underlying dynamics of the motor will be different, i.e., when systems $S^{(n)}$ and $S^{(m)}$, $i \neq j$, are excited with same input signals $\tilde{v}_{\alpha,k},\tilde{v}_{\beta,k}$, the output will differ $({i}^{(n)}_{\alpha,k},{i}^{(n)}_{\beta,k}) \neq ({i}^{(m)}_{\alpha,k},{i}^{(m)}_{\beta,k})$. Hence, by feeding $\mathcal{F}_\phi$ with $I_k^{(n)}$, the \textit{contextual filter} can distinguish among different motor instances to estimate $\omega_k^{(n)}$.

Contrary to performance indexes of conventional state-estimation techniques where the system is fixed, in cost \eqref{eq:cost} we penalize estimation errors from seen motor instances at once, such that filter \eqref{eq:contextual_filter} estimates the motor speed at instant \( k \) for any BLDC motor belonging to the class $\mathcal{S}$.

Specifically, the filter generalizes to unseen motor instances without requiring parameter updates by leveraging \textit{in-context learning} (ICL). In this paradigm, underlying patterns—such as the motor's dynamics—are inferred from a provided context, i.e., a sequence of input-output measurements.

In our application, the \textit{context} corresponds to the sequence of input-output data \( I_k^{(n)} \). Since different motors have different parameters \( \theta \), their responses to the same input signals \( \tilde{v}_{\alpha,k} \) and \( \tilde{v}_{\beta,k} \) will differ. Specifically, for two different motor instances \( S^{(n)} \) and \( S^{(m)} \), we have $(i^{(n)}_{\alpha,k}, i^{(n)}_{\beta,k}) \neq (i^{(m)}_{\alpha,k}, i^{(m)}_{\beta,k})$.
By feeding the contextual filter \( \mathcal{F}_\phi \) with \( I_k^{(n)} \), the model can distinguish between different motor instances and estimate \( \omega_k^{(n)} \).

\subsection{Transformer-based contextual filter}
The transformer architecture \cite{vaswani2017attention} is particularly suitable for in-context learning (ICL) tasks as it can efficiently process long sequences in parallel. 
Our architecture follows the design of \cite{bus24}, which is inspired by the \textit{Large Language Model} (LLM) GPT-2 \cite{radford2019language}. However, while LLMs process text tokens such as subword units, in our case, each token corresponds to a measurement tuple from the input-output sequence:  
\[
X_\kappa = \left(v^{(n)}_{\alpha,\kappa}, v^{(n)}_{\beta,\kappa}, i^{(n)}_{\alpha,\kappa}, i^{(n)}_{\beta,\kappa} \right).
\]  

For a complete description of the architecture, we refer the reader to \cite{vaswani2017attention}. Nonetheless, for completeness we here provide a synthetic description. Given an input sequence \( X = [X_1, X_2, \dots, X_{H}]^{\top} \) of length \( H \), our transformer-based filter estimates the motor speed \( \omega_k^{(n)} \) by processing the sequence through a stack of attention layers.
Each input token \( X_{\kappa} \) is projected into a \( d \)-dimensional latent space via an embedding layer $z_{\kappa} = W_{\text{emb}} X_{\kappa} + b_{\text{emb}}$.
To encode positional information, we add a learnable \textit{positional embedding}, which assigns a unique trainable vector to each position in the sequence. 
The embedded sequence is hence computed as:  
\[
e_{\kappa} = z_{\kappa} + P_{\kappa}, \quad \kappa = 1, \dots, H.
\]  
where \( P_{\kappa} \in \mathbb{R}^{d} \) is the $\kappa$-th column of the learnable embedding table \( P \in \mathbb{R}^{d \times n^{\mathrm{ctx}}} \).  
The embedded sequence $E_1 = [e_1,\ldots,e_H]$ is passed through \( l=1,\ldots, n_{\text{lyrs}} \) stacked transformer decoder blocks, each consisting of $(i)$ a first layer normalization, $(ii)$ a \textit{causal multi-head attention} that computes contextual dependencies while preventing future information leakage, $(iii)$ residual connections and layer normalization, and $(iv)$ a \textit{Feed-Forward Network} (FFN) with an additional residual connection generating the embedded sequence $E_{l+1}$.
% For each layer \( l \), the attention weights are computed as
% \begin{equation}
%     \label{eq:attention}
%     H_l = \text{softmax}\left(\frac{Q_l K_l^T}{\sqrt{d_k}} + M \right) V_l,   
% \end{equation}
% where $Q_l = E_l W_Q, K_l = E_l W_K, V_l = E_l W_V$ with $W_{(\cdot)}$ learnable weights, and \( M \) is a causal mask ensuring that the model does not attend to future steps. 
%
The output of the last layer is normalized and passed through a linear layer that provides the estimated motor speed $\hat{\omega}^{(n)}_{\phi,k}$.

\begin{remark}[Maximum allowed sequence length]
    The transformer can process sequences of shorter or equal length to the maximum context length $n^{\mathrm{ctx}}$, which is a user-defined hyperparameter. The longer $n^{\mathrm{ctx}}$, the more parameters the architecture has and the more it takes to perform one inference step (see \cite{bus24}, Section 4, Fig.~4), which must be compatible with the sampling time.
\end{remark}

% \begin{algorithm}
% \caption{Inference with Transformer-Based Contextual Filter}\label{alg:inference}
% \begin{algorithmic}[1]
% \State \textbf{Input:} Input sequence \( I^{(n)}_k = [x_1^{(n)}, x_2^{(n)}, ..., x_{n^{\text{seq}}}^{(n)}] \)
% \State \textbf{Output:} Estimated speed \( \hat{\omega}^{(n)}_{\phi,k} \)
% \State \textbf{Step 1: Embedding}
% \State Project each input token: \( z_{\kappa} = W_{\text{emb}} x_{\kappa}^{(n)} + b_{\text{emb}} \)
% \State Compute positional encoding \( P_{\kappa} \)
% \State Compute embedded sequence: \( E = [z_1 + P_1, ..., z_{n^{\text{seq}}} + P_{n^{\text{seq}}}] \)

% \State \textbf{Step 2: Transformer Layers}
% \For{each layer \( l = 1, \dots, n_{\text{lyrs}} \)}
%     \State Compute queries, keys, and values: 
%     \[
%     Q_l = E_l W_Q, \quad K_l = E_l W_K, \quad V_l = E_l W_V
%     \]
%     \State Compute causal self-attention: 
%     \[
%     H_l = \text{softmax}\left(\frac{Q_l K_l^T}{\sqrt{d_k}} + M \right) V_l
%     \]
%     \State Apply residual connection and normalization:  
%     \[
%     Z_l = \text{LayerNorm}(E_l + H_l)
%     \]
%     \State Apply feed-forward network:  
%     \[
%     F(Z_l) = \max(0, Z_l W_1 + b_1) W_2 + b_2
%     \]
%     \State Apply second residual connection and normalization:  
%     \[
%     Y_l = \text{LayerNorm}(Z_l + F(Z_l))
%     \]
% \EndFor

% \State \textbf{Step 3: Output Projection}
% \State Compute final output:  
% \[
% \hat{\omega}^{(n)}_{\phi,k} = W_{\text{out}} Y_{n^{\text{seq}}} + b_{\text{out}}
% \]

% \end{algorithmic}
% \end{algorithm}

\subsection{On the aliasing problem}\label{subsec:alias}
To perform an inference step, the transformer takes a finite amount of time \( T^{\mathrm{eval}} \).
We assume in this work a sampling time \( T_s \gg T^{\mathrm{eval}} \), so that \( T^{\mathrm{eval}} \) is negligible.

This assumption, however, introduces the problem of aliasing in the measurement of voltages and currents, depending on the operating speed of the motor. The signals \( v_\alpha, v_\beta, i_\alpha, i_\beta \) exhibit sinusoidal behavior with an electrical angular frequency \( \omega_e \) proportional to the motor's mechanical speed \( \omega \) as  $\omega_e = p \cdot \omega$, where \( p \) is the number of pole pairs. The Nyquist sampling theorem \cite{boashash2015time} states that the maximum resolvable frequency without aliasing is $\omega_{e,\max} = \frac{\pi}{T_s}$. Thus, the maximum mechanical speed that can be correctly reconstructed from sampled signals is $\omega_{\max} = \frac{\pi}{p T_s}$.
Therefore, when the motor operates at speeds above this threshold, different true speeds may produce identical observed signals, making them indistinguishable. In particular, if the motor runs at speeds exceeding \( \omega_{\max} \), the observed signals could correspond to integer multiples of the true speed \( \omega, 2\omega, ..., N \omega \), leading to ambiguity.
Nonetheless, this issue can be contained by relying on a key (mechanical) feature of the motor, namely the physical continuity of its speed. It is unlikely that the speed abruptly jumps in between integer multiples in a single timestep. This continuity constraint can indeed be leveraged by the transformer if it has access to the estimated speed from the previous timestep. While the true speed is unknown, we can feed the filter with the previous estimated speed \( \hat{\omega}^{(n)}_{\phi,k-1} \) as an additional input. Consequently, the information vector in \eqref{eq:information_vector} is modified as  
\begin{equation}\label{eq:extended_information_vector}
    \tilde{I}^{(n)}_{k} = \{ v^{(n)}_{\alpha,\kappa}, v^{(n)}_{\beta,\kappa}, i^{(n)}_{\alpha,\kappa}, i^{(n)}_{\beta,\kappa}, \hat{\omega}^{(n)}_{\phi,\kappa-1} \}_{\kappa=k-H+1}^{k},
\end{equation}
where the initial estimate \( \hat{\omega}^{(n)}_{\phi,0} \) is set to zero for the first prediction step.
As we will show, this modification allows the transformer to correctly estimate speeds beyond the aliasing limit imposed by the sampling time.
\begin{remark}[On the need of $\omega_{e,0} \leq \omega_{e,\max}$]
    The approach is effective only if the filter starts estimating while the true motor speed is below the aliasing threshold. If the initial speed exceeds this limit, the ambiguity persists from the beginning, making it impossible to recover the correct speed.
\end{remark}

\begin{remark}[Training the contextual filter]
    By modifying the information vector as in \eqref{eq:extended_information_vector}, the training process of $\mathcal{F}_\phi$ must be performed in \textit{simulation}, because the transformer has to be fed with previous speed estimates.
\end{remark}

\begin{remark}[Case with $T_s \ll T^{\mathrm{eval}}$ ]
    Handling the case where \( T_s \ll T^{\mathrm{eval}} \) would require additional measures, which we leave for future work, as the estimate \( \hat{\omega}^{(n)}_k \) at time \( k \) would be delayed by \( \lceil{{T^{\mathrm{eval}}}/{T_s}}\rceil \) steps relative to the measurements. Nonetheless for our application, diminishing $T_s$ excessively is not required, since a sampling frequency of $20-100$ Hz is sufficient to observe the speed dynamics. The corresponding $T_s = 10-50$ ms is compatible with the $T^{\mathrm{eval}} \approx1-5$ ms required for an inference step.
\end{remark}

\begin{algorithm}[!tb]
\caption{Training the contextual filter}\label{alg:cl_train}
\begin{algorithmic}[1]
\State \textbf{Input:} $\mathcal{D}$
\State \textbf{Output:} Optimal filter parameters $\phi^{\star}$
\While{$n^{\mathrm{epoch}}\leq n^{\mathrm{itr}}$}
    \For{$i=1,\ldots,b$}
        \State $\hat{\omega}^{(n)}_{{\phi},0} = 0$ 
        \For{$k = 1$ \textbf{to} $H$}
            \State $\hat{\omega}^{(n)}_{{\phi},k} = \mathcal{F}_\phi(\tilde{I}^{(n)}_k)$
        \EndFor    
        \State $\mathcal{L}^{(n)}(\phi) \gets \frac{1}{H}\sum_{k=1}^{H}\|\omega^{(n)}_k-\hat{\omega}_{{\phi},k}\|_{2}^{2}$
    \EndFor
    \State $\mathcal{L}(\phi) \gets \frac{1}{b} \sum_{i=1}^{b} \mathcal{L}^{(n)}(\phi)$
    \State $\phi \gets \phi - \eta \nabla \mathcal{L}$
\EndWhile
\State $\phi^{\star} \gets \phi$
\end{algorithmic}
\end{algorithm}

%% file: sections/experimental_results_AC.tex
\section{Experimental results}\label{sec:experimental_results}
\begin{figure}[t]
    \centering
    \includegraphics[width=\linewidth]{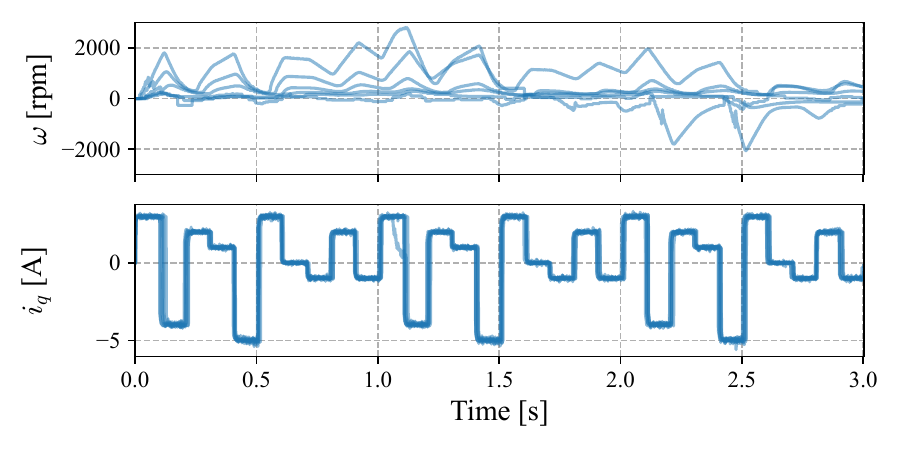}
    \caption{Speed response of different BLDC motor configurations when subjected to the same quadrature current reference profile.}
    \label{fig:OL}
\end{figure}

The experimental setup used to test the methodology is composed of\footnote{The code and the results are available in the GitHub repository \texttt{https://github.com/buswayne/in-context-bldc}.}:
\begin{itemize}
    \item a BLDC motor (Maxon EC-i 40), that has a 7 pole pairs, a rated power of 100W, rated voltage of 48V and a rated speed of 4390rpm. It is equipped with a Hall sensor, which allows us to measure the rotor position and speed for testing. The motor is rigidly clamped to a vertical surface.
    \item A bidirectional power supply, with a rated power of 6kW.
    \item A three-phase motor controller, namely a demonstration board EVSPIN32G4 (STMicroelectronics).
    \item A set of $N=6$ disks that can be mounted on the motor shaft as inertia loads. The radius $r$, mass $m$ and inertia $J$ of each disk are reported in Tab.~\ref{tab:disk_specs}. For each disk, we have a corresponding system $S^{(1)},\ldots,S^{(6)}$ each with its own dynamics.
    \item An external computing unit composed of: x64 based pc, 12th Gen Intel Core i7-12700H (4.70 GHz, 14 cores), 16Gb RAM, NVIDIA RTX 3070 Ti Laptop GPU (4Gb). The CPU can communicate with the motor controller through a UART-based network ($f_s = 1$kHz), which allows us to set the gains for the FOC algorithms and the reference speed profiles, as well as to collect experimental data. Furthermore, this unit was used to run the simulator presented in the next subsection, and to train the transformer model described in section \ref{sec:methodology}.
\end{itemize}
The disk configuration $S^{(1)}$ is used to generate simulated instances, ensuring that the simulator produces trajectories whose domain closely matches that of real motor trajectories.
The other configurations are reserved for testing. In Fig.~\ref{fig:OL}, it is displayed how the motor speed profile varies across different configurations when the same reference quadrature current $i_q$ profile is applied to the inner FOC loop of the controller. These trajectories are presented solely to highlight the diversity of configurations and the complexity of the filter design problem, but they are not used as additional training data.

For each configuration, 20 randomized closed-loop speed profiles were observed, with a sampling time $T_s = 0.01s$. According to the limits discussed in section \ref{subsec:alias}, this would imply a maximum unbiased speed $\omega_{\max}\approx 430$ rpm. As such, in these experiments, the maximum speed does not exceed this value. Specifically, each profile is a sequence of steps, whose amplitude and duration are randomly extracted in the ranges $[50, 400]$ [rpm] and $[3, 5]$ s respectively, for a total duration of 20s. Furthermore, a fixed profile was chosen across each configuration, with the following structure:
\begin{equation}
r(t) = \begin{cases}
100 &\text{if $t<5s$}  \\
200 &\text{if $t\geq5s$}\ \wedge \ t<10s\\
300 &\text{if $t\geq10s$} \wedge \ t<15s\\
150 &\text{if $t\geq15s$} \wedge \ t<20s\\
\end{cases}
\end{equation}

\begin{remark}[Sensored closed-loop experiments]
    The speed profiles from real motors are collected in a closed-loop, sensored setting. The focus of this study is solely on evaluating the filter's performance, rather than its application in sensorless control. Assessing its closed-loop performance in a fully sensorless setup would require deploying it within the actual control architecture and conducting additional experiments, which we leave for future work.
\end{remark}
\begin{figure}[]
	\centering
	\includegraphics[width=0.8\linewidth, trim=0cm 11cm 0cm 6cm,clip]{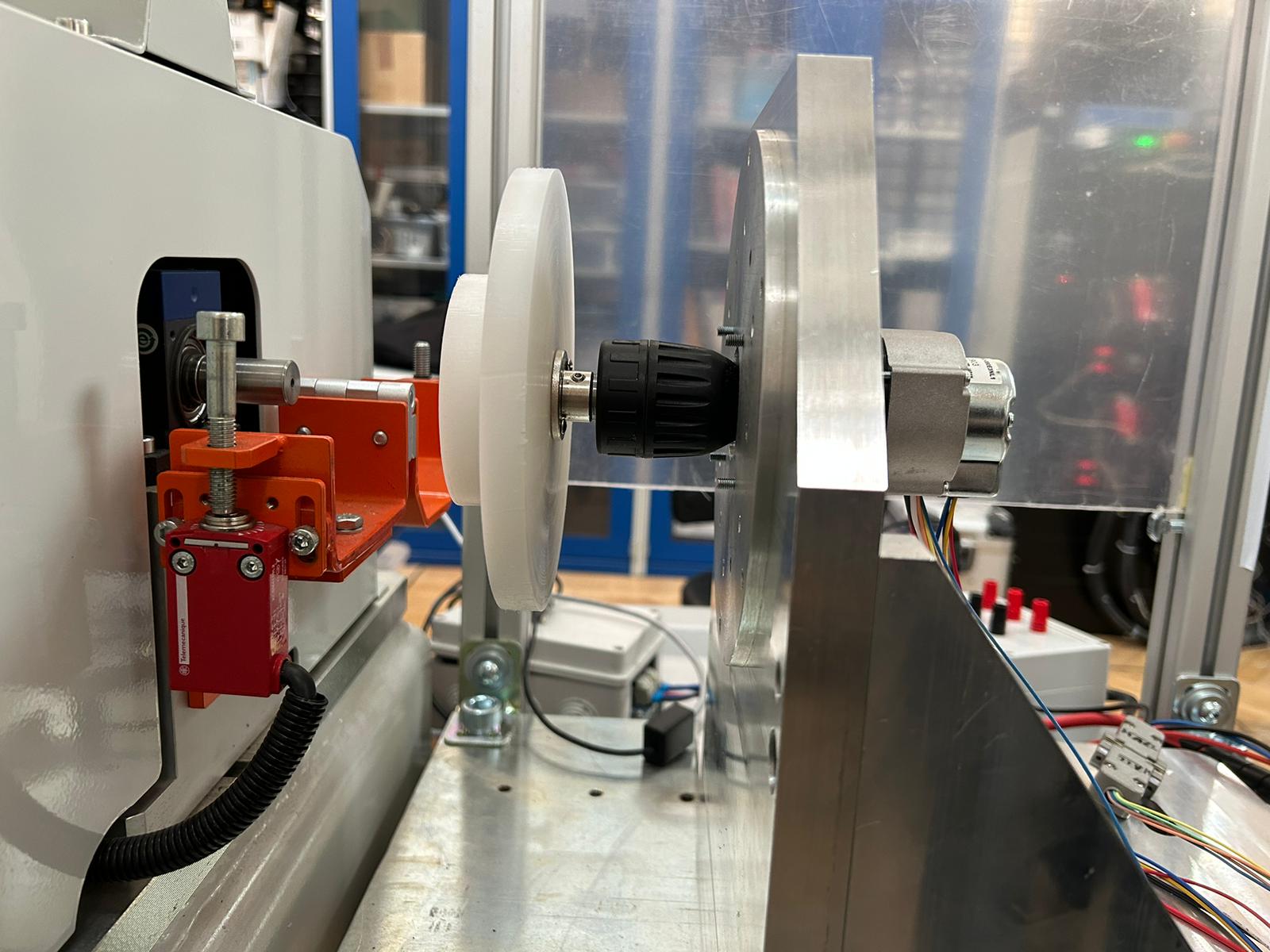}\vspace{-.2cm}
	\caption{Experimental setup for a specific configuration. Note that, the motor and the inertial load are connected through a spindle.}
	\label{fig:setup}\vspace{-.3cm}
\end{figure}

% \begin{table}[tb]
% \centering
% \addtolength{\tabcolsep}{-6pt} 
% \begin{tabular}{lcccccc}
% \hline
%          & $S^0$       & $S^1$       & $S^2$       & $S^3$      & $S^4$       & $S^5$     \\ \hline 
% $r$              & 0.05        & 0.04        & 0.06        & 0.075      & 0.08        & 0.09      \\
% $m$               & 0.114       & 0.088       & 0.175       & 0.207      & 0.212       & 0.212     \\
% $J$ & $1.425\cdot10^{-4}$ & $7.041\cdot10^{-5}$ & $3.150\cdot10^{-4}$ & $5.822\cdot10^{-4}$ & $6.784\cdot10^{-4}$ & $8.856\cdot10^{-4}$\\
% \hline
% \end{tabular}
% \caption{Load disks configurations}
% \label{tab:disk_specs}
% \addtolength{\tabcolsep}{6pt} 
% \end{table}

\begin{table}[]
\centering
\begin{tabular}{cccc}
\hline
 & $r$ [m]  &  $m$ [kg]  & $J$ [kg$\cdot m^2$]  \\
\hline
$S^{(1)}$ & 0.05  & 0.114 & $1.425\cdot10^{-4}$ \\
$S^{(2)}$ & 0.04  & 0.088 & $7.041\cdot10^{-5}$ \\      
$S^{(3)}$ & 0.06  & 0.175 & $3.150\cdot10^{-4}$ \\
$S^{(4)}$ & 0.075 & 0.207 & $5.822\cdot10^{-4}$ \\
$S^{(5)}$ & 0.08  & 0.212 & $6.784\cdot10^{-4}$ \\
$S^{(6)}$ & 0.09  & 0.212 & $8.856\cdot10^{-4}$ \\
\hline
\end{tabular}
\caption{Load disks configurations}
\label{tab:disk_specs}
\end{table}

\subsection{Simulator for the class of BLDC motors}
\begin{figure*}[!h]
    \centering
    \includegraphics[width=\linewidth]{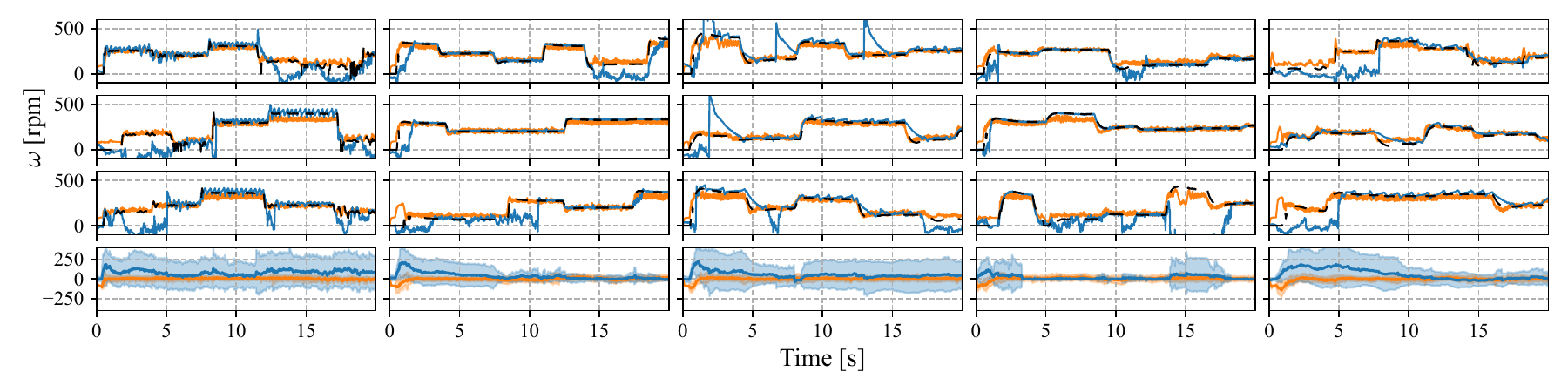}
    \caption{Examples of speed trajectories corresponding to disk configuration $S^{(2)}$ (first column) to $S^{(6)}$ (last column).  True speed (dashed-black line) is compared against the speed estimate from the contextual filter (orange) and EKFs (blue) estimations. The last row displays the estimation error and standard deviation (shaded areas) for the specific configuration averaged over the $M=15$ experiments.}
    \label{fig:gridplot}
\end{figure*}
The meta-dataset (\ref{eq:meta-set}) was generated through a BLDC motor simulator, that we use as a digital twin for the considered motor class. The main idea behind this approach is that by perturbing the parameters $\theta$ of a digital twin, we can generate trajectories that represent the behavior of motor instances belonging to the BLDC class $\mathcal{S}$.
The simulator was implemented in MATLAB/Simulink, using a modified version of a PMSM FOC model\footnote{Available at https://it.mathworks.com/help/slcontrol/ug/tune-field-oriented-controllers-using-closed-loop-pid-autotuner-block.html}.
Its components are: a BLDC motor model, a three-phase inverter, and a FOC scheme, consisting of a PI speed controller and two PI current controllers (for the direct and quadrature currents $i_d$ and $i_q$).
The parameters $\theta \in \mathbb{R}^8$ that the user is allowed to modify are the stator resistance $R_s\ [\Omega]$ and inductance $L_s\ [H]$, the maximum flux linkage $\lambda_m\ [Wb]$, the motor inertia $J_m\ [Kg\cdot m^2]$, the disk inertia $J_d\ [Kg\cdot m^2]$, the viscous damping coefficient $B_m\ [\frac{N\cdot m \cdot s}{rad}]$, the PI speed controller proportional and integral gains $K_i$ and $K_p$.
The motor and controller parameters were tuned to match the open-loop trajectories observed on the real motor, equipped with disk configuration $S^{(1)}$. Starting from data collected on the real motor, the simulator was fed the same direct and quadrature voltage; through Bayesian Optimization (BO) \cite{rasmussen2003gaussian}, the simulator parameters were fine tuned in 1000 iterations to minimize the difference between the real speed and the simulated one. The obtained parameters, defined as \emph{nominal parameters}, were then perturbed before generating each trajectory in the meta-dataset.
Specifically, each parameter is selected by multiplying their nominal quantity by a value extracted randomly from a uniform probability distribution $\mathcal{U}_{[0.5, 1.5]}$. An exception to this is the disk inertia $J_d$, whose multiplier is extracted in the range $\mathcal{U}_{[0.1, 10]}$. The list of all perturbed parameters, their nominal value, and their corresponding perturbation range can be found in Tab.~\ref{tab:sim_par}.
The number of pole pairs $p=7$ is fixed, as in the real motor.
The simulated trajectories consists in closed loop experiments with a speed profile structured as: 
\begin{equation}\label{eq:sim}
r(t) = \begin{cases}
0 &\text{if $t<0.5s$}\\
r_1 &\text{if $t\geq0.5s \ \wedge \ t<2.5s$}\\
r_2 &\text{if $t\geq2.5s \ \wedge \ t<4.5s$}\\
0 &\text{if $t\geq4.5s \ \wedge \ t<5s$}\\
\end{cases}
\end{equation}
where $r(t)$ is the reference speed, and $r_1$ and $r_2$ are two random values extracted uniformly in the range $[0, 400]$ rpm. The maximum speed of 400 [rpm] and the  sampling time $T_s = 0.01s$ were chosen in coherence with the experiments on the real motor. A total of 1000 simulated experiments were simulated, each generated with a different set of parameters $\theta^{(n)}$, sampled within the specified range.

% \begin{table}[]
% \centering
% \begin{tabular}{ccc}
% \hline
% Parameter   & Nominal value                                      & Perturbation range                          \\ \hline
% $R_s$       & $0.355 [\Omega]$                                   & $[0.1776 - 0.5327] [\Omega]$                         \\
% $L_s$       & $1.4 \cdot 10^{-3} [H]$                            & $[0.7 \cdot 10^{-3} - 2.1 \cdot 10^{-3}] [H]$   \\
% $\lambda_m$ & $1.76 \cdot 10^{-2} [Wb]$                          & $[0.88 \cdot 10^{-2} - 2.64 \cdot 10^{-2}] [Wb]$ \\
% $J_m$       & $4.4 \cdot 10^{-6} [Kg\cdot m^2]$                  & $[2.2 \cdot 10^{-6} - 6.6 \cdot 10^{-6}] [Kg\cdot m^2]$    \\
% $J_d$       & $8.73 \cdot 10^{-4} [Kg\cdot m^2]$                 & $[8.73 \cdot 10^{-5} - 8.73 \cdot 10^{-3}] [Kg\cdot m^2]$  \\
% $B_m$       & $8.3 \cdot 10^{-9} [\frac{N\cdot m \cdot s}{rad}]$ & $[4.15 \cdot 10^{-9} - 1.25 \cdot 10^{-8}] [\frac{N\cdot m \cdot s}{rad}$  \\
% $K_p$       & $0.1$                                              & $[0.05 - 0.15]$                             \\
% $K_i$       & $0.1$                                              & $[0.05 - 0.015]$                            \\ \hline
% \end{tabular}
% \caption{BLDC simulator nominal parameters and perturbation range for each}
% \label{tab:sim_par}
% \end{table}

\begin{table}[]
\centering
\caption{Perturbation range for each of the BLDC motor simulator}
\begin{tabular}{rcl}
\hline
Parameter                                & Nominal value                            &  Perturbation range                          \\ \hline
$R_s\ [\Omega]$                          & $0.355$                                  &  $[0.1776 - 0.5327]$                         \\
$L_s\ [H]$                               & $1.4 \cdot 10^{-3}$                      &  $[0.7 \cdot 10^{-3}, 2.1 \cdot 10^{-3}]$   \\
$\lambda_m\ [Wb]$                        & $1.76 \cdot 10^{-2}$                     &  $[0.88 \cdot 10^{-2}, 2.64 \cdot 10^{-2}]$ \\
$J_m\ [Kg\cdot m^2]$                     & $4.4 \cdot 10^{-6}$                      &  $[2.2 \cdot 10^{-6}, 6.6 \cdot 10^{-6}] $    \\
$J_d\ [Kg\cdot m^2]$                     & $8.73 \cdot 10^{-4}$                     &  $[8.73 \cdot 10^{-5}, 8.73 \cdot 10^{-3}] $  \\
$B_m\ [\frac{N\cdot m \cdot s}{rad}]$    & $8.3 \cdot 10^{-9}$                      & $[4.15 \cdot 10^{-9}, 1.25 \cdot 10^{-8}]$  \\
$K_p\ [\text{-}]$                        & $0.1$                                    & $[0.05, 0.15]$                             \\
$K_i\ [\text{-}]$                        & $0.1$                                    & $[0.05, 0.015]$                            \\ \hline
\end{tabular}
\label{tab:sim_par}
\end{table}

\subsection{Contextual filter}\label{subsec:filter}
The proposed contextual filter was implemented with Pytorch \cite{paszke2019pytorch}, and trained over the simulated trajectories described in the previous subsection. A list of the filter hyper-parameters can be found in \ref{tab:filter}. Compared to other contextual filters in the literature \cite{bus24}\cite{forgione2023system}, the proposed filter is notably smaller: the choice of using such a limited size is tied to the fact that the inference time must be smaller than the sampling time, which in turn must be compatible with the motor speed dynamics. In fact, this model has an average inference time of 0.00335s $\pm 0.0029$. The overall training time was of 4.34h, and the final RMSE score was of 0.0034. 

\begin{table}[]
\centering
\caption{Model hyper-parameters}
\begin{tabular}{ccccccccc}
\hline
$n_{\mathrm{params}}$ & $n_{\mathrm{layers}}$ & $n_\mathrm{heads}$ & $n_\mathrm{ctx}$ & $d$ & $n_{\mathrm{itr}}$ & batch size $b$    \\ \hline
25121        & 8            & 4           & 10        & 16  & 5000    & 128       \\ \hline
\end{tabular}
\label{tab:filter}
\end{table}

% A series of experimental data was collected on the real motor to test the performances of the filter. For each configuration, 20 randomized closed-loop speed profiles were observed. Each profile is a sequence of steps, whose amplitude and duration are randomly extracted in the ranges $[50, 400]$ [rpm] and $[3, 5]$ s respectively, for a total duration of 20s. Furthermore, a fixed profile was chosen across each configuration, with the following structure:
% \begin{equation}
% r(t) = \begin{cases}
% 100 &\text{if $t<5s$}  \\
% 200 &\text{if $t\geq5s$}\ \wedge \ t<10s\\
% 300 &\text{if $t\geq10s$} \wedge \ t<15s\\
% 150 &\text{if $t\geq15s$} \wedge \ t<20s\\
% \end{cases}
% \end{equation}

\subsection{Comparison with model-based approach (EKF)}
To evaluate the performance of our filter, $N=5$ EKFs were designed as a comparative baseline, one for each testing configuration. The calibration process of the parameters of each EKF is performed over 5 experiments, randomly selected out of the 20 experiments conducted.

Notice that this comparison favors the EKF, as it relies on data from the specific system, whereas the contextual filter operates without such prior knowledge.
The overall model uses as states the vector $[i_d, i_q, \omega, \theta_e]$:

\begin{equation*}
\begin{cases}
\dot{i}_d = -\frac{R_s}{L_s} i_d + i_q \omega + \frac{\cos{\theta_e}}{L_s} v_\alpha + \frac{\sin{\theta_e}}{L_s} v_\beta + \eta_1\\
\dot{i}_q = -\frac{Rs}{Ls} i_q + ( \frac{\lambda_m}{L_s} -i_d) \omega -  \frac{\sin{\theta_e}}{L_s} v_\alpha + \frac{\cos{\theta_e}}{L_s} v_\beta + \eta_2\\
\dot{\omega} = \frac{3}{2}\frac{p\lambda_m}{J_{tot}}i_q - \frac{B_m}{J_{tot}} \omega + \eta_3 \\
\dot{\theta}_e = p\omega + \eta_4 \\
i_\alpha = \cos{\theta_e}i_d - \sin{\theta_e}i_q + \xi_1\\
i_\beta = \sin{\theta_e}i_d - \cos{\theta_e}i_q + \xi_2 \\
i_{d,0} = \bar{i}_{d,0} + \zeta_1 \\
i_{q,0} = \bar{i}_{q,0} + \zeta_2 \\
\omega_{0} = \bar{\omega}_{0} + \zeta_3 \\
\theta_{e,0} = \bar{\theta}_{e,0} + \zeta_4
\end{cases}
\end{equation*}
where $J_{tot}= J_m + J_d$ represents the overall inertia, and $\eta = [\eta_1,\eta_2,\eta_3,\eta_4] \sim \mathcal{N}(0,Q)$, $\xi = [\xi_1,\xi_2] \sim \mathcal{N}(0,R)$, and $\zeta = [\zeta_1,\zeta_2,\zeta_3,\zeta_4] \sim \mathcal{N}(0,P_0)$ are white noises affecting the process, measurement, and initial state respectively. The model implicitly uses the Park transform to convert the inputs and outputs from the $\alpha \beta$ framework to the direct and quadrature one.
$R_s, L_s, \lambda_m, J_{tot}, \text{ and } B_m$ were obtained from the additional experiments. 
The covariance matrices $R$, $Q$, and $P_0$ used by the EKF are defined as:
% \begin{align}
%     R = \mathrm{diag}(1,1), \quad Q = \mathrm{diag}(q_1,q_2,q_3,q_4), \\
%     \quad P_0 = \mathrm{diag}(10^{-6},10^{-6},10^{-6},p_0),
% \end{align}

\begin{equation}
    \begin{aligned}
        &R = \mathrm{diag}(1,1), \quad Q = \mathrm{diag}(q_1,q_2,q_3,q_4), \\
        &\hphantom{R = {}} P_0 = \mathrm{diag}(10^{-6},10^{-6},10^{-6},p_0)
    \end{aligned}
\end{equation}

% \begin{equation}
% R = \begin{bmatrix}
% 1 & 0 \\
% 0 & 1 
% \end{bmatrix} \qquad
% Q = \begin{bmatrix}
% q_1 & 0 & 0 & 0 \\
% 0 & q_2 & 0 & 0 \\
% 0 & 0 & q_3 & 0 \\
% 0 & 0 & 0 & q_4 
% \end{bmatrix}    
% \end{equation}
%
% \begin{equation}
% P_0 = \begin{bmatrix}
% 10^{-6} & 0 & 0 & 0 \\
% 0 & 10^{-6} & 0 & 0 \\
% 0 & 0 & 10^{-6} & 0 \\
% 0 & 0 & 0 & p_0 
% \end{bmatrix}    
% \end{equation}
%
% Where $R$ is the covariance of the measurement noise, $Q$ is the covariance of the process noise, and $P_0$ is the initial state covariance.
% $q_1,q_2,q_3,q_4,\text{ and } p_0$ were chosen to minimize the estimation error of $\omega$. We use $x_0 = [0,0,0,0]$ as the initial state, and we assume that at the start of our experiment, the motor is still and no current is flowing, although the exact initial angle is unknown, hence the structure of $P_0$.
where $q_1,q_2,q_3,q_4,\text{ and } p_0$ are tunable parameters chosen to minimize the estimation error of $\omega$. We set the initial state to $x_0 = [\bar{i}_{d,0},  \bar{i}_{q,0},  \bar{\omega}_{0}, \bar{\theta}_{e,0}] = [0,0,0,0]$. Therefore, we assume that at the beginning of our experiment, the motor is still and no current is flowing. Nonetheless, as the exact initial angle is actually unknown, we reflect this through the chosen structure of $P_0$ (where we show less trust in the initial condition for the angle).

Based on these choices, note that the EKF uses the same signals as the contextual filters, i.e., the voltages $v_{\alpha}, v_{\beta}$, and the currents $i_{\alpha}, i_{\beta}$.
Finally, the EKFs calibrated based on 5 sets of experimental data were tested to estimate the speed over the remaining (15) datasets and, hence, their respective disk configurations.

\subsection{Analysis of the results}
\begin{figure}[!ht]
    \centering
    \includegraphics[width=\linewidth]{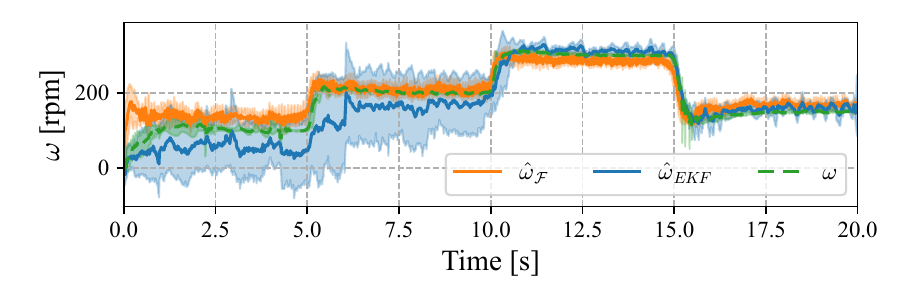}
    \caption{Average estimation along the fixed reference experiments.}
    \label{fig:multiplot}
\end{figure}
\begin{figure}[!ht]
    \centering
    \includegraphics[width=\linewidth]{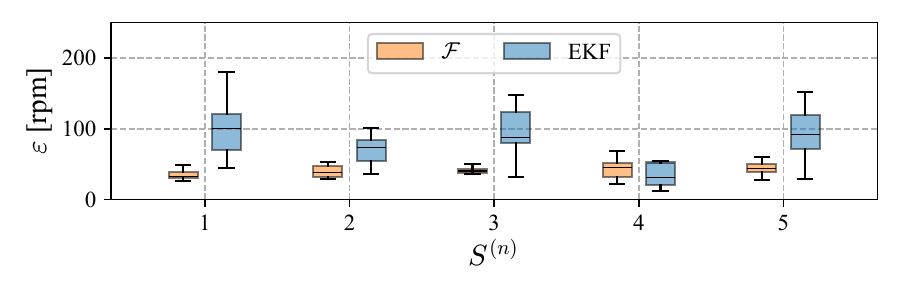}
    \caption{Boxplot of the estimation error $\varepsilon$ for the specific testing configurations averaged over time and the experiments.}
    \label{fig:boxplot}
\end{figure}
\begin{figure*}[ht]
    \centering
    \includegraphics[width=\linewidth]{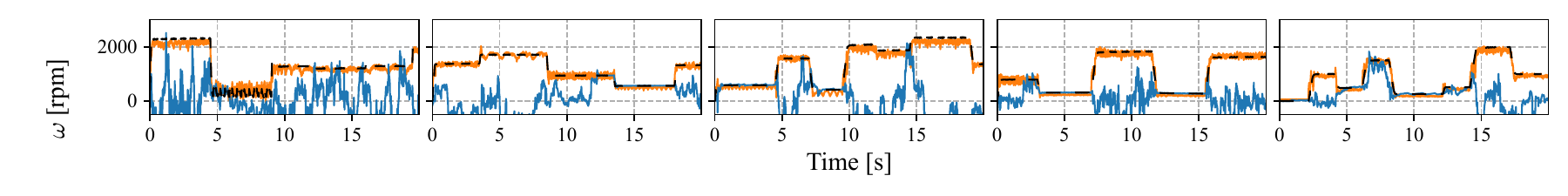}
    \caption{Examples of speed trajectories with $\omega \geq \omega_{\max}$, corresponding to configurations $S^{(2)}, \ldots, S^{(6)}$. The estimated speed with $\mathcal{F}_\phi$ (orange) can reconstruct the true speed (dashed black line), while the EKFs estimate (blue) fails.}
    \label{fig:high}
\end{figure*}
In Fig. \ref{fig:gridplot}, we can see the true speed of the motor compared to the estimates provided by the contextual filter and the configuration-tailored EKF. In the last row of the figure, the average estimation error for the two methods is reported. By looking at the plot, we can %appreciate how 
see that both methods %it converges 
converge to an unbiased estimate of the true speed. However, the standard deviation of the contextual filter %remains 
is significantly smaller %with respect to 
than the one %of 
achieved with the EKF, as it is more clearly %visualized 
visible in the boxplot of Fig.~\ref{fig:boxplot}. 
%Indeed, by comparing the in-context estimator and EKF on some example trajectories, we can see that %while 
%the estimate of the contextual filter is always centered around the true value, while the EKF errors drastically (and suddenly) increase in correspondence of abrupt changes. 
In some segments, the EKF (see 2nd row, 4th column) is almost perfect, highlighting how much the performance of EKF depends on the operating condition and how sensitive they are to their changes due to nonlinear effects. Note also that, at the beginning of these experiments, the contextual filter has a non-negligible transient error, because the context is not sufficiently long to distinguish the particular motor instance. Nonetheless, it is clear that the more data the in-context estimator is fed with, the more its performance gets better.

Similar conclusions can be drawn from Fig. \ref{fig:multiplot} in which the average true and estimated speed across the five test configurations is shown for the fixed profile experiments. After an initial transient, the contextual filter is able to track the true speed across all tests, while the EKFs struggle to track it correctly for the first half of the experiment. This result highlights the resilience of the contextual filter to changes in the motor configuration. %It is clear how the contextual filter is more consistent in it estimations. 
This consideration is corroborated by the comparison between the estimation RMSE obtained with the in-context estimator and EKF for each disk configuration shown in Fig. \ref{fig:boxplot}. Indeed, these results highligh %highlighting 
how the contextual filter outperforms the EKF, in terms of average RMSE and its variance, in all but one configuration (in which they show similar results). The (poorer) results attained with EKF come with the additional burden of a more involved design of the filter, with EKF requiring model identification and parameter tuning from real-world data to be designed. %One must also consider the fact that, for each configuration, model identification and parameter tuning from real-world data was required to design an appropriate EKF, while 
Instead, the contextual filter only required simulated data and needed no modification to work on a specific configuration.

\subsection{Beyond aliasing}
As mentioned in Section \ref{subsec:alias}, the choice of a sampling time $T_s>>T^{\mathrm{eval}}$ can introduce an aliasing problem. While the recursive architecture of our filter is able to mitigate the issue, this is not the case for the EKF. To demonstrate this fact, a second filter was trained without changing the hyper-parameters shown in Tab.~\ref{tab:filter} on simulated trajectories equivalent to the ones described in \eqref{eq:sim} in which the coefficients $r_1$ and $r_2$ are extracted in the range $[0,4000]$ rpm. Similarly, we collected data from the real motor by providing randomized speed profiles such as the ones described in Section \ref{subsec:filter}, but in which the step amplitude is extracted in the range $[250, 2500]$ rpm. An exemplification of the profile for each disk configuration can be seen in Fig. \ref{fig:high}. As clear from these results, the contextual filter can estimate the speed of the real motor correctly. On the other hand,the EKF can sometimes track the actual motor speed at a lower velocities, but fails to do so as the velocity increases above the limit discussed in Sec.~\ref{sec:methodology} ($\omega_{\max}\approx 430$ rpm).

%% file: sections/conclusions.tex
\section{Conclusion and Future Works}\label{sec:conclusions}
% In this work, we presented an experimental validation of the approach proposed in \cite{bus24} on an industrial applications, namely the speed estimation in sensorless brushless motors.
% %
% Our findings confirm that with in-context learning, we can design relying on simulated instances an unique state estimator that when deployed does not require any further fine-tuning regardless the specific motor instance of the considered BLDC class. 

% Future research will focus on the deployment of the transformer-based architecture in embedded systems for online estimation of the motor's speed and the expansion of the class of motors that the filter can deal with, and practical validation of contextual controllers.

In this work, we presented an experimental validation of the approach proposed in \cite{bus24} on an industrial application, namely speed estimation in sensorless brushless motors.
Our results confirm that, through in-context learning, we can design a state estimator for a whole class of BLDCs using simulated instances. When deployed, this estimator requires no further fine-tuning, regardless of the specific motor instance within the considered class, achieving superior performance in comparison to model-based methods (EKF).

Future research will focus on ($i$) deploying the transformer-based architecture in embedded systems for real-time motor speed estimation, ($ii$) expanding the class of motors the filter can handle, and ($iii$) providing an experimental validation for the dual-problem of designing a contextual controller \cite{busetto2024one}.